\documentstyle{article}

\begin{document}

\title{Bulk viscosity of interacting strange quark matter }
\author{Zheng Xiaoping$^{1,2}$ Yang Shuhua$^1$ Li Jiarong$^2$ 
\\
{\small 1 Department of Physics, Huazhong Normal University}\\
\small 2 Institute of Particle Physics, Huazhong University}
\date{}
\maketitle
\begin{abstract}
The  quarks are regarded as  quasiparticles, which acquire an
effective mass generated by the interaction with the other quarks
of the dense system.   The bulk viscosity is re-expressed in the
case of interacting strange quark matter. We find that the
viscosity is a few$\sim$tens times larger than that of
non-interacting quark gas due to the small correction of medium
effect to the equation of state. The limiting spins of strange
stars will rise
so significantly that the pulsars with shorter period may exist.
\end{abstract}

 As the hypothesis compact objects, strange stars have not been proven by
 observation yet. Probing strange stars and then identifying them are  significant challenges in
nuclear physics and astrophysics. It has been shown in a series of
papers, that the emission of gravitational radiation due to r-mode
instabilities in hot, young neutron stars severely limits the
rotation period of these stars\cite{1,2,3,4,5,6}. Whereas Madsen
found that the r-mode instability does not play any role in a
young strange star contrary to young neutron stars\cite{7}. The
author reasonably concluded that finding a young pulsar with a
rotation period below 5-10 milliseconds would be a strong
indication of the existence of strange stars. In another
article\cite{8}, the author also discussed the evolution of
pulsars as strange stars due to r-mode instabilities in rapidly
rotating stars. The results revealed that strange stars, perhaps
better than neutron stars, explained the apparent lack of very
rapid pulsars with frequencies above 642Hz sine existing data are
consistent with pulsars being strange stars under some
circumstances.All above involve the viscosities of strange quark
matter contained in the interior of these stars. This is because
of viscosity damping the r-modes. As known, the shear viscosity of
quark matter\cite{7,9} is roughly comparable to  that of neutron
star matter for the parameter range of our interest where the bulk
viscosity of quark matter, which damps the r-modes, is decisive
for the localization of the instability, differs  from neutron
star matter. The bulk viscosity of strange quark matter mainly
arises from the non-leptonic weak interactions\cite{10}
\begin{equation}
u+d\leftrightarrow s+u \end{equation}

The importance of dissipation due to the reaction (1) was first
stressed by Wang and Lu\cite{11} in the case of neutron stars with
quark cores. Afterward, Sawyer\cite{12} expressed the damping as a
function of temperature and oscillation frequency. Furthermore,
Madsen in 1992\cite{13} considered the non-linearity of the
reaction rate and corrected the factor error of the rate
in\cite{12} to obtain the exact formula of strange matter. All
these works showed a huge bulk viscosity of strange matter
relative to nuclear matter. Possible implications of a pulsar
being strange star will result from the discrepancy of bulk
viscosities between strange quark matter and normal nuclear
matter.

However, up to now, the considerations of the viscosity are based
on the equation of state(EOS) described as a free Fermi gas of
quarks at zero temperature. In fact, the coupling among quarks (i.
e. so-called medium effect) due to strong interaction is very
important for quark-gluon plasma(QGP). Many investigations devoted
to the field. We here hope to uncover the medium effect on the
bulk viscosity of strange matter. As known, one of the most
important medium effects are effective masses generated by the
interaction of the quarks with the system. Schertler  with his
colleagues  had studied the contribution of effective mass effect
to EOS of strange matter\cite{14} and applied the EOS to
astrophysical aspect for understanding of strange star's
structure\cite{15}, where the authors applied the idea of an ideal
gas of quasiparticles with effective masses to the case of cold
strange matter.

We follow ref\cite{14}: At finite chemical potential($\mu _{i}$),
the effective (quark) mass of quasiparticle is obtained from zero
momentum limit of the quark dispersion following from the hard
dense loop(HDL) approximation\cite{16} of the quark self energy.
In the case of a vanishing current quark mass, as assumed for up
and down quarks, the effective mass is given as follows\cite{17}

\begin{equation}
m_{q}^{*2}=\frac{g^{2}\mu _{q}^{2}}{6\pi ^{2}}
\end{equation}
For strange quarks due to non-vanishing current quark masses, we
incorporate the current mass into the effective one\cite{18}

\begin{equation}
m_{s}^{*}=\frac{m_{s}}{2}+\sqrt{\frac{m_{s}^{2}}{4}+\frac{g^{2}\mu _{s}^{2}}{%
6\pi ^{2}}}
\end{equation}
where $g$ is coupling constant of strong interaction. An upper
limit for the coupling constant is given by $g=7.7$ in\cite{14}.
As was used in \cite{14}, we here take $g$ as a parameter ranging
from 0 to 5. Furthermore we replace the dispersion relation of the
collective quarks branch by $\varepsilon _{{\bf k}}=\sqrt{{\bf
k}^{2}+m^{*2}}$

Evidently, transport coefficients depend mainly on  EOS for given
reaction(1)\cite{11,12,13}. Medium effects affect the EOS, then
the viscosity will be changed. Although effective mass effect on
structure(or EOS) of a strange star is small\cite{14}, the
viscosity is very sensitive to the EOS. In the following, we will
show that this is indeed the case.

We continue to adopt the quasiparticle picture to treat an
ensemble of quasiparticles as a free, degenerate Fermi gas of
particles with a total energy

\begin{equation}
H_{\rm{eff}}=\sum_{i=1}^{d}\sum_{{\bf k}}\sqrt{{\bf k}^{2}+m_{i}^{*2}(\mu )%
}n_{{\bf k}}+B^{*}V
\end{equation}
Due to the $\mu $-dependence of $m^{*}(\mu )$ the function $E^{*}(\mu )$ $%
=B^{*}(\mu )V$ defines a necessary energy counterterm in order to
maintain thermodynamic self-consistency\cite{14}. $d$ denotes the
degree of degeneracy and $\sum_{k}n_{k}=N$ is particle number of
system. Thus the partition function has

\begin{eqnarray}
Z
=e^{-\beta E^{*}}\{\prod_{{\bf k}}[1+e^{-\beta (\varepsilon _{k}-\mu
)}\}^{d}
\end{eqnarray}
where $\beta =\frac{1}{T},T$ is the temperature. The grand
thermodynamic potential $\Omega$ in a volume is easily derived.


We immediately get the pressure

\begin{eqnarray}
P &=&-\frac{\partial \Omega }{\partial V}|_{\mu } \\
&=&\frac{d}{48\pi ^{2}}\left[ \mu k_{{\rm F}}(2\mu ^{2}-5m^{*2})+3m^{*4}\ln
{k_{{\rm F}}+\mu  \overwithdelims() m^{*}}%
\right] -B^{*}(\mu )  \nonumber
\end{eqnarray}
We also obtain other static quantities such as particle number density and
energy density according to thermodynamic relations

\begin{equation}
\rho (\mu )=\frac{{\rm d}P}{{\rm d}\mu }|_{m^{*}}
\end{equation}

\begin{equation}
\epsilon (\mu )=\mu \frac{{\rm d}P}{{\rm d}\mu }|_{m*}-P(\mu )
\end{equation}
To maintain these fundamental relations above, we should require
constraint condition\cite{14}

\begin{equation}
{\partial P \overwithdelims() \partial m^{*}}%
_{\mu }=0
\end{equation}
 $B^{*}$ will be uniquely defined following from the constraint
equation(9).

Therefore the particle and energy densities is explicitly expressed as

\begin{equation}
\rho =\frac{d}{6\pi ^{2}}k_{{\rm F}}^{3}
\end{equation}

\begin{equation}
\epsilon =\frac{d}{16\pi ^{2}}\left[ \mu k_{{\rm F}}(2\mu
^{2}-m^{*2})-m^{*4}\ln
{k_{{\rm F}}+\mu  \overwithdelims() m^{*}}%
\right] +B^{*}(\mu )
\end{equation}

This, combining (6)with(10)and(11), is  the EOS of the strange
matter derived by Schertler, Greiner and Thoma in quasiparticle
description, together with MIT bag constant $B_{0}$, in $\beta
$-equilibrium\cite{14}


Now we start with  the bulk viscosity of SQM in quasiparticle
description. We follow the method of derivation of the bulk
viscosity which was presented by Wang and Lu\cite{11} and then
developed by Madsen\cite{13}.

We still keep the Wang and Lu's notation for convenience, defining $v$, the
volume per unit mass and $n_{i}$, the quark numbers per unit mass. Assume
the volume vibrates around the equilibrium volume $v_0$, with the period $\tau$
and the perturbation amplitude $\Delta v$.

When the amplitude is small, i. e. $\frac{\triangle v}{v_{0}}$ $\ll
1 $, we express the pressure $P(t)$ through a expansion near equilibrium $%
P_{0} $

\begin{equation}
P(t)=P_{0}+%
{\partial P \overwithdelims() \partial v}%
_{0}\delta v+%
{\partial P \overwithdelims() \partial n_{d}}%
_{0}\delta n_{d}+%
{\partial P \overwithdelims() \partial n_{s}}%
_{0}\delta n_{s}
\end{equation}
According to the reaction (1), we shall have

\begin{equation}
\delta n_{d}=-\delta n_{s}\equiv \int_{0}^{t}\frac{{\rm d}n_{d}}{{\rm d}t}
\end{equation}

If the net rate of reaction (1) $\frac{{\rm d}n_{d}}{{\rm d}t}$
and EOS in equilibrium state were given, the mean dissipation rate
of the vibrational energy per unit mass can be obtain from
following expression

\begin{equation}
{{\rm d}w \overwithdelims() {\rm d}t}%
_{{\rm ave}}=-\frac{1}{\tau }\int_{0}^{\tau }P(t)\frac{{\rm d}v}{{\rm d}t}%
{\rm d}t
\end{equation}
It has been certified by Wang and Lu\cite{11} that the first two
terms in (13) don't contribute to the energy dissipation rate
(15). This conclusion is independent of concrete $P_{0}$. The
derivative of the third and fourth terms in (12) can be obtained
from thermodynamical relation

\begin{equation}
\frac{\partial P}{\partial n_{i}}=-\frac{\partial \mu _{i}}{\partial v}
\end{equation}
We need slightly change Eq(10) in place of $\rho $ with $n_{i}$

\begin{equation}
n_{i}=\frac{1}{\pi }k_{{\rm F}i}^{3}v=\frac{1}{\pi }(\mu
_{i}^{2}-m_{i}^{*2})^{\frac{3}{2}}v
\end{equation}
It immediately reads

\begin{equation}
\left( \frac{\partial \mu _{i}}{\partial v}\right) _{0}=-\frac{k_{{\rm F}%
i}^{2}}{3v_{0}C_{i}}
\end{equation}

\begin{equation}
{\partial \mu _{i} \overwithdelims() \partial n_{i}}%
_{0}=\frac{\pi ^{2}}{3v_{0}k_{{\rm F}i}C_{i}}
\end{equation}
where $C_{i}=\mu _{i}-m_{i}^{*}\frac{\partial m^{*}}{\partial \mu _{i}}$.
One gets

\begin{equation}
{\partial P \overwithdelims() \partial n_{d}}%
_{0}\delta n_{d}+%
{\partial P \overwithdelims() \partial n_{s}}%
_{0}\delta n_{s}=\frac{1}{3v_{0}}\left( \frac{k_{{\rm F}d}^{2}}{C_{d}}-\frac{%
k_{{\rm F}s}^{2}}{C_{s}}\right) \int_{0}^{t}\frac{{\rm d}n_{d}}{{\rm d}t}
\end{equation}

We continue to use the formula of the reactional rate adopted by
Madsen\cite{13}

\begin{equation}
\frac{{\rm d}n_{d}}{{\rm d}t}\approx \frac{16}{5\pi ^{2}}G_{F}^{2}\sin
^{2}\theta _{c}\cos ^{2}\theta _{c}\mu _{d}^{5}{\rm \delta }\mu [{\rm \delta
}\mu ^{2}+4\pi ^{2}T^2]v_{0}
\end{equation}
with

\begin{equation}
\frac{16}{5\pi ^{2}}G_{F}^{2}\sin ^{2}\theta _{c}\cos ^{2}\theta
_{c}=6.76\times 10^{-26}{\rm MeV}^{-4}
\end{equation}
The chemical potential difference ${\rm \delta }\mu $ can be
derived like the pressure in Eq(12)

\begin{equation}
{\rm \delta }\mu (t)=%
{\partial {\rm \delta }\mu  \overwithdelims() \partial v}%
_{0}\delta v+%
{\partial {\rm \delta }\mu  \overwithdelims() \partial n_{d}}%
_{0}\delta n_{d}+%
{\partial {\rm \delta }\mu  \overwithdelims() \partial n_{s}}%
_{0}\delta n_{s}
\end{equation}
Substituting Eqs (17) and (18) into above formula, we have

\begin{equation}
{\rm \delta }\mu =\frac{1}{3}\left( \frac{k_{{\rm F}d}^{2}}{C_{d}}-\frac{k_{%
{\rm F}s}^{2}}{C_{s}}\right) \frac{\triangle v}{v}\sin
{2\pi t \overwithdelims() \tau }%
-\frac{\pi ^{2}}{3}\frac{1}{v}\left( \frac{1}{k_{{\rm F}d}C_{d}}-\frac{1}{k_{%
{\rm F}s}C_{s}}\right) \int_{0}^{t}\frac{{\rm d}n_{d}}{{\rm d}t}
\end{equation}

The definition of the bulk viscosity is from\cite{12,13}

Finally, it reads

\begin{equation}
\zeta =\frac{1}{\pi v_{0}}%
{v_{0} \overwithdelims() \triangle v}%
\frac{1}{3}\left( \frac{k_{{\rm F}d}^{2}}{C_{d}}-\frac{k_{{\rm F}s}^{2}}{%
C_{s}}\right) \int_{0}^{\tau }{\rm d}t\left[ \int_{0}^{t}\frac{{\rm d}n_{d}}{%
{\rm d}t}\right] \cos
{2\pi t \overwithdelims() \tau }%
\end{equation}
Our derivations are made in quasiparticle description. We find
from these equations that $B^{*}$ modifying the EOS due to the
medium effect don't contributes to the viscosity, but the
effective mass  play an important role on it. The  prefactors in
(19), (23) and (24) differ from that in \cite{7,8,13}. They depend
not only the chemical potential $\mu $ but also coupling constant
$g$ due to $m^{*}$-dependence of $k_{{\rm F}i}$ and $C_{i}$. When
taking $g=0$, the mass of u,d-quark vanishes,
 we have $k_{{\rm F}d}=\mu _{d},k_{{\rm F}
s}=(\mu _{d}^{2}-m_{s}^{2})^{\frac{1}{2}},C_{d}=C_{s}=\mu _{d}$,
and then the results above go back to Madsen's. It has been
realized that the medium effect of EOS expressed in Eq(10) and
(11) is small under the study\cite {14}, but the small
contribution to EOS greatly influences on the dynamical quantity
(24). Hence the viscosity $\zeta $ is sensitive to the mass
$m^{*}$ although the EOS is weakly mass-dependence under the
assumption of weak coupling.


Combining Eq(20) with Eq (22), the quantities ${\rm \delta }\mu $, $\frac{{\rm d%
}n_{d}}{{\rm d}t}$ can be calculated in principle, but a general solution
can't appear analytically because of strong coupling of the two equations.
We will give the solution in the high temperature approximation and then
the numerical solution for general case similar to that made by Madsen.


 When the
temperature is high enough, i. e.$2\pi T\gg {\rm \delta }\mu $,
the cubic term ${\rm \delta }\mu ^{3}$ in Eq(20) can be neglected.
The equations (20) and (22) are easily solved analytically. With
Madsen's expression, the bulk viscosity reads

\begin{equation}
\zeta =\frac{\alpha ^{*}T^{2}}{\omega ^{2}+\beta ^{*}T^{4}}\left[ 1-\left[
1-\exp \left( -\beta ^{*\frac{1}{2}}T^{2}\tau \right) \right] \frac{2\beta
^{*\frac{1}{2}}T^{2}/2}{\omega ^{2}+\beta ^{*}T^{4}}\right]
\end{equation}
The symbol $*$ is used to distinguish the our result from Madsen's. Here
$\omega =\frac{2\pi }{\tau }$, and

\begin{equation}
\alpha ^{*}=9.39\times 10^{22}\mu _{d}^{5}\left( \frac{k_{{\rm F}d}^{2}}{%
C_{d}}-\frac{k_{{\rm F}s}^{2}}{C_{s}}\right) ^{2}{\rm (gcm}^{-1}{\rm s}%
^{-1})
\end{equation}

\begin{equation}
\beta ^{*}=7.11\times 14^{-4}\left[ \frac{\mu _{d}^{5}}{2}\left( \frac{1}{k_{%
{\rm F}d}C_{d}}-\frac{1}{k_{{\rm F}s}C_{s}}\right) \right] ^{2}{\rm (s}^{-2})
\end{equation}

These factors depend on not only the current mass $m_{s}$ but also
the coupling $g$ when the chemical potential is given. $\alpha
,\beta $ in \cite {13} is independent of the coupling constant.
This is because of inclusion of medium dependent mass effect in
our study.


As Madsen has pointed out, the dominating term in rate(20) is
proportional to $ \delta \mu ^{3}$ at low temperature (2$\pi T\ll
\delta \mu $). Therefore, in general Eqs(20), (22) and (24) must
be solved numerically. The results of such calculations for the
given parameters $g,m_{s}$ and $\mu $ are shown in figure 1. The
viscosity is much larger than that of non-interacting quark gas,
where the maximum increase of the  viscosity  is  larger by about
two orders of magnitude. To understand deeply  the medium effect,
we also give the $g$-dependence of the curves of the viscosity
versus $\Delta v/v, T$, shown in Figs2, and 3. In principle, the
viscosity become larger with increasing $g$. Analogous to Madsen's
result, the linear effect, i.e. the constant viscosity in Fig 2,
 dominates at tiny vibrations while the viscosity is quickly enlarged when there exist
 the somewhat larger vibrations. Fig 3 shows that the effect of interacting medium
 on the viscosity is rather huge except in the extreme high temperature limit, over a few
 $\times 10^9$K. To understand the observed pulsar's data,  the viscosity for the parameter
 range of interest is
 a few $\times 10^5\sim$a few$\times 10^7$K  because Madsen
 find that the viscosity in the range play an important role if   pulsars is
  strange stars\cite{8}. Because the medium effect can enlarges the bulk viscosity
  of SQM, the critical spin frequency curves in frequency-temperature space
  (see the figures in \cite{8}) will be modified. Fig 3 in \cite{8} shows that
  a 2-flavor color superconducting phase(2SC) seems marginally ruled out by the
  observation data, but we can expect
that 2SC quark matter stars may be safe absolutely if the medium effect is included.

In summary, the bulk viscosity of SQM including medium effect is derived. The influence
of interacting medium on the viscosity shouldn't be negligible. We find that then
the maximal discrepancy of about two orders of magnitude appears. We
may expect that the limiting angular velocity curves in frequency-temperature space
will be change due to the larger bulk viscosity, which will be discussed in detail
in another study.

 This work was supported in part by the National Natural Science Foundation of China
 under Grant No.10175026.

Figure captions: the viscosity $\zeta$ in gcm$^{-1}$s$^{-1}$\\
Fig1. A comparison of the our bulk viscosity(solid curves $g$=4) with non-interacting SQM(dotted curves),
$m_s$=80MeV , $\mu_d$=470MeV, $\tau$=0.001s
 From bottom to top: $T=10^{-5},10^{-4},10^{-3},10^0,10^{-2},10^{-1}$MeV\\
Fig2. Bulk viscosity as function of perturbation for
$ m_s$=150MeV , $\mu_d$=470MeV, $\tau$=0.001s,
      $T=10^{-4}$MeV.
From bottom to top:  g=0,1,2,3,4,5\\
Fig3. Bulk viscosity as function of temperature for
$ m_s$=150MeV , $\mu_d$=470MeV, $\tau$=0.001s,
      ${\Delta v\over v}=10^{-4}$.
From bottom to top:  g=0,1,2,3,4,5


\begin{thebibliography}{99}
\bibitem{1}N. Andersson, Astrophys. J.502(1998)708.
\bibitem{2}J. L. Friedman and S. M. Morsink, Astrophys. J. 502(1998)714.
\bibitem{3}L. Lindblom, B. J. Owen and S. M. Morsink, Phys. Rev.
Lett. 80(1998)4843.
\bibitem{4} B. J. Owen et al., Phys. Rev. D58(1998)084020.
\bibitem{5}N. Andersson, K. D. Kokkotas and B. F. Schutz, Astrophys. J. 510(1999)846.
\bibitem{6}N. Andersson, K. D. Kokkotas and N. Stergioulas, Astrophys. J. 516(1999)307.
\bibitem{7}J. Madsen, Phys. Rev. Lett. 81(1998)3311.
\bibitem{8}J. Madsen, Phys. Rev. Lett. 85(2000)10.
\bibitem{9} H. Heiselberg and C. J. Pethick, Phys. Rev. D48(1993)2916
\bibitem{10}H. Heiselberg, Phys. Scr. 46(1992)485; J. Madsen,
Phys. Rev. D47(1993)325.
\bibitem{11}  Q.D. Wang, T. Lu, Phys. Lett B148(1984)211.
\bibitem{12}  R. F. Sawyer, Phys. Lett. B233(1989)412.
\bibitem{13}  J. Madsen, Phys. Rev. D46(1992)3290.
\bibitem{14}  K. Schertler, C. Greiner and M. H. Thoma,Nucl.Phys. A616(1997)659.
\bibitem{15}  K. Schertler, C. Greiner J. Schaffner-Bielich and M. H. Thoma,Nucl.Phys. A677(1997)463.
\bibitem{16}  C. Manuel, Phys. Rev. D53(1996)5866.
\bibitem{17}   H. Vija and M. H. Thoma, Phys. Lett. B342(1995)212.
\bibitem{18}  R. D. Pissarski, Nucl. Phys. A498(1989)423c; J.-P. Blaizot and
J.-Y. Ollitrault, Phys. Rev. D48(1993)1390.

\end{thebibliography}
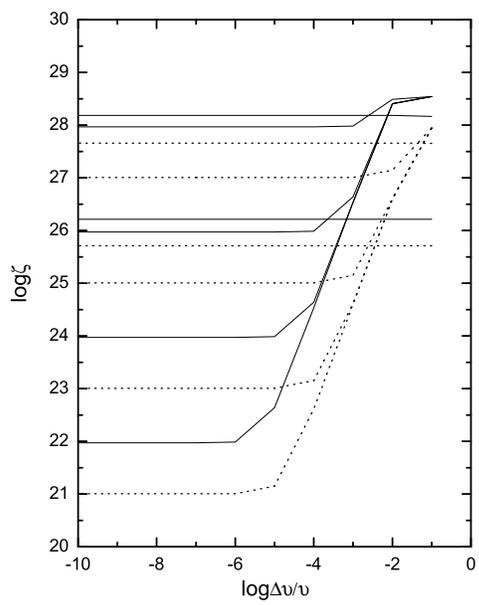
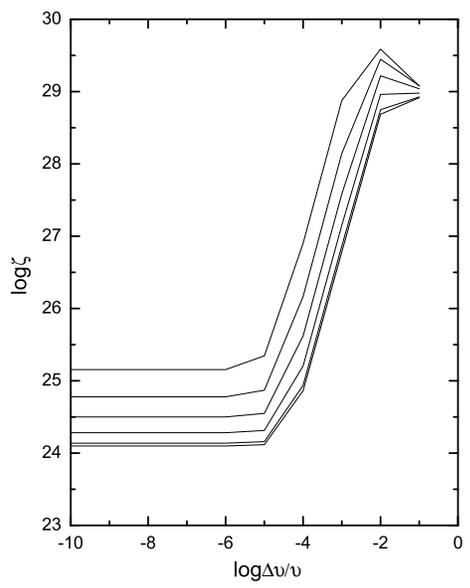
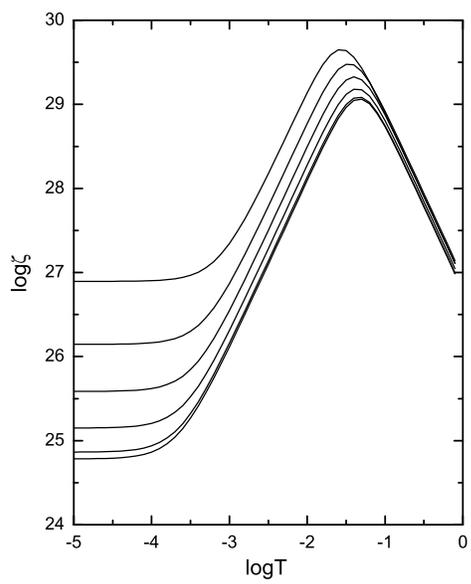
\end{document}